\documentclass[a4paper,fleqn,usenatbib]{mnras}

\usepackage{newtxtext,newtxmath}

\usepackage[T1]{fontenc}
\usepackage{ae,aecompl,multirow}

\usepackage{graphicx}	
\usepackage{amsmath}	
\usepackage{amssymb}	
\usepackage{subfig}
\usepackage{xcolor}

\title[Spectral Turnover of PKS\;2155-304]{Establishing the Spectral Turnover of Blazar PKS\;2155-304 as an Outcome of Radiative Losses}

\author[J. Sitha et al.]{
Sitha K. Jagan$^{1}$\thanks{E-mail: sithajagan@gmail.com},
S. Sahayanathan$^{2}$,
R. Misra$^{3}$,
C. D. Ravikumar$^{1}$
and K. Jeena$^{4}$
\\
$^{1}$Department of Physics, University of Calicut, Malappuram-673635, India\\
$^{2}$Astrophysical Sciences Division, Bhabha Atomic Research Centre, Mumbai - 400085, India\\
$^{3}$ Inter-University Center for Astronomy and Astrophysics, Post Bag 4, Ganeshkhind, Pune-411007, India\\
$^{4}$Department of Physics, Providence Women's College, Malaparamba, Calicut-673009, India
}

\date{Accepted XXX. Received YYY; in original form ZZZ}

\pubyear{2017}

\begin{document}
\label{firstpage}
\pagerange{\pageref{firstpage}--\pageref{lastpage}}
\maketitle

\begin{abstract}

The broad band optical/UV and X-ray spectra of blazars have been often modeled 
as synchrotron component arising from a broken power-law distribution of electrons. 
A broken power-law distribution is expected, since the high energy electrons undergo 
radiative losses effectively. The change in the energy index should then be $\approx1$ 
and corresponds to a spectral index difference of 0.5. 
However, one of the long outstanding problems has been that the observed index change is significantly different. On the other hand, recent high quality observations of blazars suggest that their local 
spectra may not be a power-law, instead have a slight curvature and often represented 
by a log parabola model. Using XMM-Newton observations spanning over 12 years for the 
BL Lac PKS\;2155-304, we show that the optical/UV and X-ray spectra can be well represented 
by a broken log parabola model. Further, we show that such a spectrum can indicate  
the energy dependence of the electron escape timescale from the main acceleration zone.
This novel approach, besides addressing the observed difference in the photon spectral
indices, also tries to explain the spectral turn over 
in far UV/soft X-rays as a result of the radiative losses.

\end{abstract}

\begin{keywords}
	galaxies: active -- BL Lacertae objects: individual: PKS\,2155-304 -- acceleration of particles -- diffusion -- X-rays: galaxies
\end{keywords}


\section{Introduction}

The low energy synchrotron spectral component of blazars is reproduced by a broken power-law function 
suggesting the underlying electron distribution to be of broken power-law in shape \citep{1994ApJS...95..371S}. 
A power-law electron distribution can be achieved
under Fermi acceleration process, where the electrons gain energy while being scattered by magnetic turbulent
structures embedded in the jet or by crossing a shock front \citep{2007Ap&SS.309..119R}. Subsequently, the synchrotron
losses modify the accelerated electron distribution into a broken power-law with 
the indices differing by unity \citep{1962SvA.....6..317K,1987MNRAS.225..335H}. 
However, the observed difference between the low and high energy
synchrotron spectral components of blazars cannot be perceived within the synchrotron cooling
interpretation of a power-law electron distribution \citep{2012ApJ...753..154M,2010ApJ...715L..16M}.

Around the peak of the synchrotron component, the blazar spectra deviate considerably from a power-law
with the spectra showing smooth curvature. 
In many cases, the spectra at the peak are
well reproduced by a log-parabola function suggesting the underlying electron distribution to be a
log-parabola as well \citep{2004A&A...413..489M}. The log parabola function however, is successful 
in explaining only a narrow band of the spectrum falling around the peak but fails to explain the 
spectrum over a broad energy range e.g., optical--X-ray energy bands \citep{2004A&A...413..489M}. 
Alternatively, the broadband synchrotron component of blazars is often fitted with a smooth broken power-law
function or a power-law with an exponential cut off \citep{2017ApJ...836...83S}. 

PKS\,2155-304 is a BL Lac class of blazars located at a redshift $z=0.116$.
Its synchrotron component peaks at UV energies \citep{2016ApJ...831..142M}
and the broadband spectrum reflects a smooth broken power-law function \citep{1999ApJ...521..552C,2005A&A...442..895A}. 
An extensive study of PKS\,2155-304 was carried out by \emph{XMM-Newton} at optical/UV (180-600 nm) and 
X-ray (0.15-12 keV) energy bands \citep{2017ApJ...850..209G,2014MNRAS.444.3647B} on different epochs spanning 
more than a decade. 
The broadband spectrum obtained through joint analysis of \emph{NuSTAR} and \emph{XMM-Newton} observations,  
supplemented with \emph{Fermi} observations at gamma-ray energies, could be reproduced satisfactorily by a 
synchrotron and synchrotron self Compton emission models due to a broken power-law electron distribution \citep{2016ApJ...831..142M}.
However, narrow band X-ray analysis showed significant deviation from a power-law spectra \citep{2017ApJ...850..209G}. 
\cite{2008A&A...478..395M} performed a detailed X-ray analysis of the source using \emph{BeppoSAX}, \emph{XMM-Newton} 
and \emph{Swift} observations from 1996 to 2007. They showed that the X-ray spectrum is well reproduced by a 
log-parabola function with the peak of the spectral energy distribution indicating a positive correlation with the spectral curvature.
Nevertheless, the log-parabola model did not succeed well in explaining the combined optical/UV and the X-ray spectrum by \emph{XMM-Newton} \citep{2014MNRAS.444.3647B}.

In this work, we perform a detailed examination of \emph{XMM-Newton} observations of PKS\,2155-304 at 
optical/UV and X-ray energies. 
The optical/UV and X-ray data from 19 November 2000 to 28 April 2012 are 
analysed and the source spectra are obtained.
The composite spectrum is then fitted with 
a broken log-parabola function. The rationale behind this being, 
the radiative cooling of a log-parabola electron distribution eventually modify it into a broken log-parabola
distribution. 
We also show that such an electron distribution can be achieved when the electron escape timescale from the main acceleration
region is energy dependent. Further, the analytical model developed under this scenario can fit the 
observed optical/UV and X-ray well and in turn, can hint the particle diffusion processes in AGN jets.

\section{Observations and Data Reduction}
We selected twenty XMM Newton archival data of PKS 2155-304, starting from 19 November 2000 to 28 April 2012,
such that they have at least one simultaneous Optical/UV exposures with X-ray.
For X-ray data, we used only the European Photon Imaging Camera (EPIC)-pn data and EPIC MOS data were avoided due to their low sensitivity, quantum efficiency and chances of pile up. This data were reduced using XMM-Newton Science Analysis System (SAS Version 14.0) following standard procedures. The calibrated photon event files for the pn camera were produced using the command \emph{epchain}. For pn data processing, both single and double events (PATTERN $\le4$) of good quality (FLAG = 0) in the energy range 0.2-10 keV were considered. In the 10-12 keV energy range, a ``Good Time Interval (GTI)" event list was produced by studying the light curve and fixing a threshold rate to omit background particle flaring. 

The source spectrum was obtained using a circular region of size $40^{\prime\prime}$ around the source. The background was estimated using two circular regions of similar size located away from the source on the same source CCD chip. For nine observations, \emph{epatplot} indicated significant pile up and the source spectrum was extracted using an annular ring of inner radius $10^{\prime\prime}$ and an outer radius between $38^{\prime\prime}$ and $40^{\prime\prime}$, within the CCD chip. 

The Optical Monitor (OM) observations during the selected epochs were reprocessed with SAS pipeline \emph{omichain}. The optical/UV data contain a significant galactic contamination which can manifest as large systematic error. To investigate this, we fitted the optical/UV data by a simple power-law function with addition of appropriate systematic error on data, required for better fit statistics. Four observations were discarded as they contained less than three optical/UV filter exposures. Similarly, two other observations demand a large systematic error ($\gtrsim$ 30\%) for a better statistics and hence they were also omitted in the present study. On an average, we found adding 3\% systematic error to the rest of 14 observations can result in better fit statistics. The details of these observations are given in Table \ref{tab:obs}.

\begin{table*}
  \centering
  \begin{minipage}{160mm}
	  {\scriptsize
\begin{tabular}{|c|l|c|c|c|c|c|c|c|}
\hline
\multirow{2}{*}{Obs.ID} &\multirow{2}{*}{Exposure (s)} & \multicolumn{7}{|c|}{Flux(${10}^{\text{-}11}$erg/cm$^{2}$/s)} \\ \cline{3-9}
 & & X-Ray & UVW2 & UVM2 & UVW1 & U & B & V \\
\hline
${0124930501}^*$ & 	104868 & ${7.534}^{+0.036}_{-0.037}$ & ${0.782}^{+0.039}_{-0.039}$ & $-$ & $-$ & ${2.171}^{+0.110}_{-0.111}$ & ${1.004}^{+0.051}_{-0.051}$ & ${4.419}^{+0.291}_{-0.278}$\\
0124930601 & 114675 & ${5.161}^{+0.013}_{-0.012}$ & ${0.733}^{+0.036}_{-0.037}$ & ${1.844}^{+0.092}_{-0.092}$ & ${1.256}^{+0.062}_{-0.063}$ & ${1.989}^{+0.101}_{-0.099}$ & ${0.955}^{+0.048}_{-0.048}$ & ${3.607}^{+0.198}_{-0.194}$\\
0158960101 & 27159 &  ${4.492}^{+0.017}_{-0.018}$ & $-$ & ${2.68}^{+0.136}_{-0.135}$ & ${1.841}^{+0.091}_{-0.092}$ & ${2.909}^{+0.148}_{-0.148}$  & $-$ & $-$  \\
0158960901 & 28919 &  ${4.849}^{+0.0162}_{-0.0162}$ & $-$ & $-$ & $-$ & ${2.981}^{+0.152}_{-0.149}$ & ${1.38}^{+0.070}_{-0.069}$ & ${5.674}^{+0.349}_{-0.337}$\\
0158961001 & 40419 & ${6.714}^{+0.017}_{-0.017}$  & ${1.05}^{+0.052}_{-0.052}$ & ${2.599}^{+0.129}_{-0.130}$ & ${1.851}^{+0.092}_{-0.092}$ & $-$ &${1.341}^{+0.067}_{-0.067}$ & $-$ \\
${0158961301}^*$ & 60415 & ${10.617}^{+0.033}_{-0.033}$  & ${2.032}^{+0.102}_{-0.101}$ & ${5.603}^{+0.280}_{-0.281}$ & ${3.778}^{+0.187}_{-0.188}$ & ${5.989}^{+0.303}_{-0.302}$ & ${2.86}^{+0.144}_{-0.144}$  & ${14.695}^{+0.906}_{-0.880}$  \\
${0158961401}^*$ & 64814 & ${4.506}^{+0.023}_{-0.022}$  & ${1.277}^{+0.064}_{-0.064}$ & ${3.561}^{+0.180}_{-0.179}$ & ${2.622}^{+0.131}_{-0.129}$ & ${4.299}^{+0.221}_{-0.218}$ & ${2.12}^{+0.108}_{-0.108}$ & ${10.729}^{+0.722}_{-0.687}$ \\
0411780101 & 101012 & ${6.02}^{+0.015}_{-0.022}$ & ${1.855}^{+0.093}_{-0.092}$ & ${5.012}^{+0.248}_{-0.249}$ & ${3.394}^{+0.167}_{-0.168}$ & ${5.652}^{+0.279}_{-0.280}$ & ${2.749}^{+0.136}_{-0.136}$ & ${14.685}^{+0.738}_{-0.747}$  \\
${0411780201}^*$ & 67911 & ${13.178}^{+0.036}_{-0.035}$ & ${2.143}^{+0.107}_{-0.107}$ & ${5.715}^{+0.284}_{-0.285}$ & ${4.049}^{+0.201}_{-0.199}$ & ${6.872}^{+0.342}_{-0.344}$ & ${3.123}^{+0.156}_{-0.155}$ & ${15.842}^{+0.887}_{-0.871}$ \\
${0411780301}^*$ & 61216 & ${16.125}^{+0.042}_{-0.043}$ & ${1.588}^{+0.079}_{-0.079}$ & ${4.104}^{+0.205}_{-0.205}$	& ${2.793}^{+0.138}_{-0.139}$ & ${4.138}^{+0.206}_{-0.207}$ & ${1.947}^{+0.098}_{-0.097}$ & ${7.901}^{+0.426}_{-0.418}$ \\
${0411780401}^*$ & 64820 & ${8.811}^{+0.024}_{-0.025}$ & ${1.739}^{+0.087}_{-0.087}$ & ${4.589}^{+0.229}_{-0.229}$ & ${3.181}^{+0.158}_{-0.157}$ & ${5.162}^{+0.259}_{-0.259}$ & ${2.377}^{+0.119}_{-0.119}$ & ${10.209}^{+0.588}_{-0.577}$  \\
0411780501 & 74298 & ${5.254}^{+0.019}_{-0.019}$ & ${0.959}^{+0.048}_{-0.049}$	& ${2.461}^{+0.124}_{-0.124}$ & ${1.642}^{+0.081}_{-0.082}$ & ${2.493}^{+0.128}_{-0.128}$ & ${1.182}^{+0.061}_{-0.059}$ & ${5.109}^{+0.315}_{-0.301}$ \\
${0411780601}^*$ &	63818 & ${7.523}^{+0.026}_{-0.026}$ & ${1.022}^{+0.053}_{-0.050}$ & ${2.545}^{+0.127}_{-0.128}$ & ${1.792}^{+0.089}_{-0.089}$ & ${2.844}^{+0.143}_{-0.143}$ & ${1.354}^{+0.068}_{-0.068}$ & ${5.551}^{+0.318}_{-0.310}$  \\
0411780701 & 68735 & ${1.466}^{+0.007}_{-0.007}$ & ${0.452}^{+0.023}_{-0.024}$ & ${1.254}^{+0.064}_{-0.064}$ & ${0.841}^{+0.042}_{-0.042}$ & ${1.369}^{+0.071}_{-0.070}$ & ${0.668}^{+0.035}_{-0.034}$  & ${3.414}^{+0.238}_{-0.199}$ \\
\hline
\end{tabular}
\caption{Observation details of PKS 2155-304 with XMM-Newton and the best fit X-ray and optical/UV fluxes. Pile up were noticed for the observation IDs with $^*$.}
\label{tab:obs}
}
\end{minipage}
\end{table*}

\section{Broken Log-parabola Model}\label{sec:blp}
The optical/UV and the X-ray data of PKS\,2155-304 for the selected epochs were fitted with X-ray Spectral Fitting Pacakage (XSpec) using 
user-defined (local) and the inbuilt models \citep{1996ASPC..101...17A}. The X-ray absorption due to Galactic neutral hydrogen in the direction of PKS\,2155-304 was estimated by fixing the hydrogen column density to $1.71\times10^{20}$ cm$^{-2}$ \citep{2014MNRAS.444.3647B}. The optical/UV data were corrected for galactic reddening using the model UVRED by setting the parameter $E_{B-V} = 0.019$ \citep{1979MNRAS.187P..73S,2011ApJ...737..103S}. Similar to earlier works, the X-ray spectra were found to significantly deviate from a power-law and were better represented by a log parabola, but failed to explain the optical/UV data \citep{2017ApJ...850..209G,2014MNRAS.444.3647B}.

In order to develop a consistent model capable of fitting both the optical/UV and X-ray data, we considered a 
scenario where a log parabola electron distribution is losing its energy under a synchrotron emission 
process \citep{2004A&A...413..489M}. For a small curvature, the radiative losses will steepen the index 
by $\sim$ 1 (Appendix \ref{app:lp}).
If the escape of electrons from the main emission region is also considered, then the electron distribution will
transform into a broken log parabola distribution with break occuring at an energy where the electron cooling timescale 
is equal to the escape timescale. 
The synchrotron spectrum resulting from such an electron distribution will again be 
a broken log parabola with the index differing by $\approx 0.5$ \citep{RAA4107}.

\begin{figure}
\includegraphics[width=0.45\textwidth]{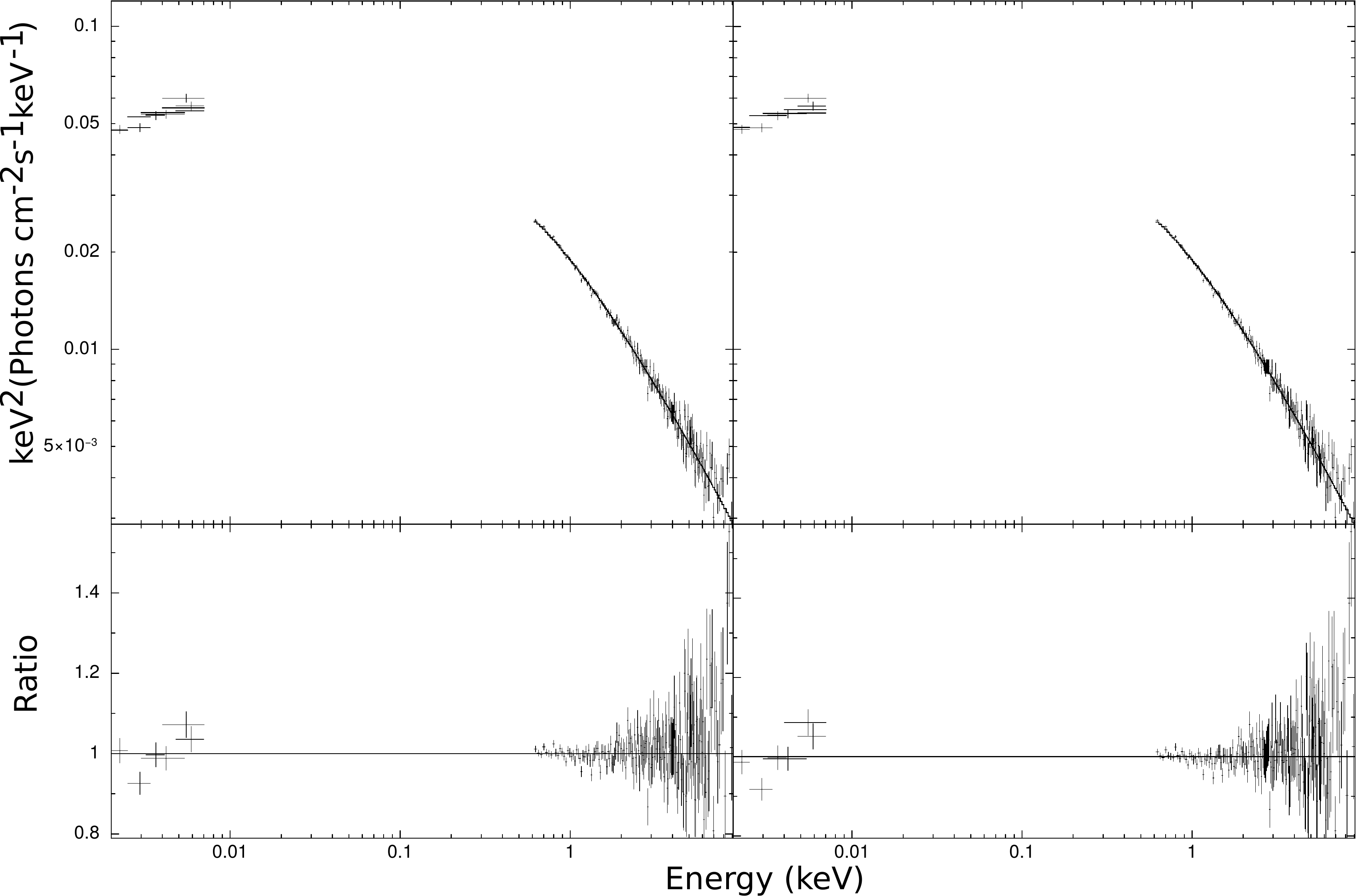}
\caption{Unfolded optical/UV and X-ray spectrum along with best fit models for Observation ID 0411780501. The models are
broken log-parabola (left) and synchrotron emission with energy dependent electron escape (right).\label{fig:fit}} 
\end{figure}

To diagnose whether this interpretation is capable of explaining the broadband distribution of PKS\,2155-304,
we performed a joint fitting of optical/UV and X-ray (0.6-10 keV) data using a broken log parabola function defined by

\begin{align}\label{eq:logpb}
	F(\epsilon)\propto\left\{
		\begin{array}{ll}
			\left(\frac{\epsilon}{\epsilon_b}\right)^{-\alpha+\Delta-\beta \rm{log}(\epsilon/\epsilon_b)}, & \rm{for}\; \frac{\epsilon}{\epsilon_b} \leq 1 \\ 
\left(\frac{\epsilon}{\epsilon_b}\right)^{-\alpha-\beta \rm{log}(\epsilon/\epsilon_b)}, & \rm{for}\; \frac{\epsilon}{\epsilon_b}>1 
		\end{array}
		\right.
\end{align}
Here, $\epsilon$ is the photon energy, $\epsilon_b$ is the break energy\footnote{Here, $\epsilon_b$ is the energy
at which the two log parabola functions assume the same value and not the peak of the log parabola.}, $\alpha$ is the index, $\Delta$ is
the difference between the indices and $\beta$ is the curvature parameter. 
Motivated by the synchrotron cooling interpretation of a log parabola electron distribution, we fixed $\Delta$ at $0.5$
and obtained the best fit parameters. The fit results are given in  Table \ref{tab:fitres}. We also provide fit statistics, assuming a power-law model with a log parabola tail (PLLP) following \citep{2014MNRAS.444.3647B}, for comparison. In Fig. \ref{fig:fit} (left) we show the spectral fit for the observation ID 0411780501. 
The best fit X-ray, UV and optical fluxes are also given in Table \ref{tab:obs}. The X-ray spectra corresponding to the observation IDs 0158961401, 0411780101 and 0411780201 show significant negative curvature suggesting a plausible contribution of Compton spectral component. Hence, for these observations the model parameters cannot be constrained effectively. Presence of Compton contamination was evident in 
most of the observations with the high energy data deviating considerably from the best fit model (fig. \ref{fig:fit}).

A Spearman rank correlation analysis between the flux and the fitted parameters was performed to investigate their dependence. 
We found no significant correlation between these quantities. However, a weak negative correlation between the curvature parameter $\beta$ and the optical/UV flux can be seen with the lowest null hypothesis probability $P_{rs}=0.03$ and corresponding rank correlation coefficient $r_s=-0.62$. This correlation is not very significant and hence we speculate that the spectral shape does not have any implication on the observed flux level rather may depend on dynamics which are unrelated to the flux variation. A plausible scenario can be the spectral curvature being dependent on the electron diffusive processes which may not have a direct association with the flux variation.

\begin{table*}
  \centering
  \begin{minipage}{160mm}
{\scriptsize
\begin{tabular}{|c|c|c|c|c|c|c|c|c|c|}
\hline
\multirow{3}{*}{Obs.ID} & \multicolumn{4}{c|}{Broken log-parabola} & \multicolumn{1}{c|}{PLLP} & \multicolumn{4}{c|}{Energy dependent escape}  \\ \cline{2-10} 
                    & $\alpha$  & $\beta$     &$\epsilon_b$     & $\chi^{2}/dof$ & $\chi^{2}/dof$   & $\psi$     & $\beta^\prime$     & $\epsilon_b^\prime$     & $\chi^{2}/dof$    \\ \hline
0124930501 & ${2.49}^{+0.05}_{-0.03}$ & ${0.083}^{+0.012}_{-0.011}$ & ${0.15}^{+0.07}_{-0.05}$ & ${199.6}/{167}$ & $324.14/167$      &  $16.06^{+4.60}_{-2.69}$       & $0.12_{-0.02}^{+0.02}$    & $0.13_{-0.04}^{+0.07}$  & $200.69/167$   \\
0124930601 & ${2.35}^{+0.04}_{-0.09}$ & ${ 0.137}^{+0.011}_{-0.004}$ & ${0.05}^{+0.01}_{-0.02}$ & ${302.7}/{204}$  &  $300.41/204$   &  $8.03_{-0.90}^{+1.03}$        & $0.21_{-0.02}^{+0.02}$    & $0.04_{-0.01}^{+0.01}$  &  $282.26/204$  \\
0158960101 & ${2.78}^{+0.04}_{-0.06}$ & ${0.071}^{+0.010}_{-0.010}$ & ${0.39}^{+0.21}_{-0.15}$ & ${191.9}/{166}$  & $355.02/166$     &  $31.29_{-8.62}^{+10.12}$      & $0.08_{-0.02}^{+0.03}$    & $0.39_{-0.17}^{+0.21}$  & $191.70/166$  \\
0158960901 & ${2.72}^{+0.05}_{-0.15}$ & ${0.135}^{+0.020}_{-0.011}$ & ${0.21}^{+0.12}_{-0.11}$ &  ${176.6}/{167}$  & $278.26/167$    &  $14.65_{-2.57}^{+3.23}$       & $0.17_{-0.02}^{+0.02}$    & $0.19_{-0.07}^{+0.12}$  & $179.76/167$   \\
0158961001 & ${2.45}^{+0.08}_{-0.07}$ & ${0.148}^{+0.008}_{-0.011}$ & ${0.07}^{+0.03}_{-0.02}$ & ${285.0}/{193}$ & $396.31/193$     &  $10.05_{-1.05}^{+1.66}$        & $0.20_{-0.02}^{+0.01}$    & $0.08_{-0.01}^{+0.03}$  & $287.91/193$  \\
0158961301 & ${2.48}^{+0.03}_{-0.06}$ & ${0.073}^{+0.011}_{-0.006}$ & ${0.07}^{+0.02}_{-0.02}$ & ${293.8}/{196}$  & $480.18/196$     &  $18.27_{-2.72}^{+3.75}$       & $0.11_{-0.02}^{+0.01}$    & $0.06_{-0.01}^{+0.02}$  & $293.70/196$  \\
0411780301 & ${2.33}^{+0.06}_{-0.04}$ & ${0.105}^{+0.005}_{-0.010}$ & ${0.06}^{+0.02}_{-0.01}$ &  ${233.6}/{209}$ & $382.12/209$     &  $10.99_{-1.37}^{+1.62}$       & $0.16_{-0.01}^{+0.01}$    & $0.06_{-0.01}^{+0.02}$  & $234.47/209$ \\
0411780401 & ${2.60}^{+0.07}_{-0.04}$ & ${0.114}^{+0.006}_{-0.016}$ & ${0.13}^{+0.07}_{-0.02}$ & ${275.9}/{200}$  & $500.10/200$   &  $15.86_{-1.97}^{+2.21}$       & $0.14_{-0.01}^{+0.01}$    & $0.13_{-0.03}^{+0.04}$  & $283.91/200$  \\
0411780501 & ${2.47}^{+0.06}_{-0.07}$ & ${0.123}^{+0.011}_{-0.009}$ & ${0.07}^{+0.03}_{-0.02}$ &  ${217.5}/{174}$  & $327.97/174$   &  $11.76_{-1.63}^{+1.87}$       & $0.17_{-0.02}^{+0.02}$    & $0.07_{-0.02}^{+0.02}$  & $225.30/174$  \\
0411780601 & ${2.29}^{+0.06}_{-0.05}$ & ${0.080}^{+0.008}_{-0.009}$ & ${0.03}^{+0.01}_{-0.01}$ & ${231.0}/{195}$  & $261.58/195$   &  $12.40_{-2.06}^{+2.62}$       & $0.13_{-0.02}^{+0.02}$    & $0.03_{-0.01}^{+0.01}$  & $229.30/195$  \\
0411780701 & ${2.66}^{+0.06}_{-0.06}$ & ${0.105}^{+0.010}_{-0.011}$ & ${0.12}^{+0.05}_{-0.03}$ & ${233.8}/{154}$  & $361.85/154$   &  $18.45_{-2.80}^{+3.40}$       & $0.13_{-0.02}^{+0.02}$    & $0.12_{-0.03}^{+0.05}$  & $239.36/154$  \\
\hline
\end{tabular}
\caption{Fit results of optical/UV and X-ray spectrum using broken log-parabola model and the synchrotron emission with energy dependent electron escape from the main acceleration region. In column 6 we provide the fit statistics for a power-law + log-parabola model (PLLP) following \protect\cite{2014MNRAS.444.3647B} for comparison and the fit parameters are similar to the values reported.\label{tab:fitres}}}
\end{minipage}
\end{table*}

\section{Electron Escape time-scale and Spectral Curvature}\label{sec:engtesc} To interpret the spectral curvature and the broken log parabola representation of the observed optical/UV and X-ray spectra of PKS\,2155-304, we considered a scenario where the electrons are accelerated at a shock front and escape into the downstream region where they lose their energy through radiative processes. Consistently, we label the region around the shock front as Acceleration Region (AR) and the downstream region as Emission Region (ER). In general, the steady state non-thermal electron distribution $n(\gamma)$ generated by an acceleration process which is balanced by the escape rate can be expressed as \citep[e.g.][]{1962SvA.....6..317K}

\begin{align}
\frac{d}{d\gamma} \left[\left(\frac{\gamma}{\tau_a}-\zeta\gamma^2\right)n(\gamma)\right] + \frac{n (\gamma)}{\tau_e} = Q_o\delta(\gamma-\gamma_0)
\label{equ:kinetic}
\end{align}
Here, we assume a mono energetic injection into AR at energy $\gamma_0$, $\gamma$ being the dimensionless electron energy and, 
$\tau_a$ and $\tau_e$ define the characteristic acceleration and escape timescales, respectively. The radiative loss rate encountered by the electrons in AR is determined by $\zeta \gamma^2$. Expressing $\zeta$ in terms of maximum Lorentz factor $\gamma_{max}$ attained by the electron, $\zeta=(\gamma_{max}\tau_a)^{-1}$, equation (\ref{equ:kinetic}) can be reduced, for $\gamma_o<\gamma\ll\gamma_{max}$, to
\begin{align}
	\frac{d}{d\gamma} \left[\frac{\gamma}{\tau_a}n(\gamma)\right] = -\frac{n(\gamma)}{\tau_e} 
\end{align}
This can be rewritten for an energy independent $\tau_a$ as
\begin{align}
\frac {d ~\hbox{ln} S(\gamma)}{d ~\hbox {ln} \gamma} = -\xi
\label{equ:kineticred}
\end{align}
where, $S(\gamma) = \gamma\, n(\gamma)$ is the energy density and $\xi = \tau_a/\tau_e$.
If we assume $\tau_e$ also to be energy independent, then the resulting electron distribution will be a powerlaw 
$n (\gamma) \propto \gamma^{-p}$ with index $p = \xi+1$. On the other hand, if we consider the escape timescale
to be energy dependent such that $\xi$ takes a form $\xi = \xi_0\, \gamma^{\beta^\prime}$, then equation (\ref{equ:kineticred}) will have a solution
\begin{align}\label{eq:arsoln}
	S(\gamma) \propto \rm{exp}\left(-\frac{\xi_0}{\beta^\prime}\,\gamma^{\beta^\prime}\right)
\end{align}

Similarly, the steady state non-thermal electron distribution $N(\gamma)$ in ER can be expressed as
\begin{align}\label{eq:erkinetic}
	-\frac{d}{d\gamma} \left[\eta\,\gamma^2 \,N(\gamma)\right] + \frac{N (\gamma)}{t_e}= \frac{n(\gamma)}{\tau_e(\gamma)}
\end{align}
where, $\eta \gamma^2$ is the radiative loss rate in ER and $t_e$ is the
characteristic escape timescale from ER. For energy independent $t_e$, 
solution to equation (\ref{eq:erkinetic}) will be of the form
\begin{align}\label{eq:erpart}
	N(\gamma)\propto\left\{
		\begin{array}{ll}
			\gamma^{\beta^\prime-1}\textrm{exp}\left(-\frac{\xi_0}{\beta^\prime}\gamma^{\beta^\prime}\right), & \rm{for}\; \gamma\ll\gamma_b \\ 
			\gamma^{-2}\textrm{exp}\left(-\frac{\xi_0}{\beta^\prime}\gamma^{\beta^\prime}\right), & \rm{for}\; \gamma\gg\gamma_b 
		\end{array}
		\right.
\end{align}
where, $\gamma_b$ corresponds to the energy for which the cooling time scale equals to $t_e$. The synchrotron photon flux
due to this electron distribution will be \citep{RAA4107}
\begin{align}\label{eq:erphot}
	F_{\rm syn}(\epsilon)\propto\left\{
		\begin{array}{ll}
\epsilon^{-1+\beta^\prime/2}\,\rm{exp}\left(-\psi \epsilon^{\beta^\prime/2}\right), & \rm{for}\; \frac{\epsilon}{\epsilon_b^\prime}\ll1 \\ 
\epsilon^{-3/2}\,\rm{exp}\left(-\psi \epsilon^{\beta^\prime/2}\right), & \rm{for}\; \frac{\epsilon}{\epsilon_b^\prime}\gg1 
		\end{array}
		\right.
\end{align}
where, $\psi$ is a parameter relating the observed photon frequency with electron energy and $\epsilon_b^\prime$ is the 
emitted photon energy corresponding to electron energy $\gamma_b$. For $\beta^\prime\ll1$, 
equation (\ref{eq:erphot}) will be equivalent to equation (\ref{eq:logpb}) with the parameters related as 
\begin{align}
  \beta^\prime \approx\frac{3.5\,\beta}{2\alpha-3},\;  \psi \approx \frac{(2\alpha-3)^2}{3.5\,\beta}
\end{align}

The observed optical/UV and X-ray data of PKS\,2155-304 for the selected epochs 
were fitted using the synchrotron spectrum from ER given by (\ref{eq:erphot}). 
Similar to the broken log parabola function, we found that this physical model can also fit
observations well and in Table \ref{tab:fitres}, we provide the fit results.
In Fig. \ref{fig:fit} (right) we show the spectral fit corresponding to this model for the observation ID 0411780501.
This study thereby suggests the observed curvature in the X-ray spectrum of blazars
may indicate energy dependence of the escape time scale. This in turn will depend on the
electron diffusion in the jet medium and hence can provide information regarding the magnetic field 
structure of blazar jets.

\section{Discussion}
We show that the combined optical/UV and X-ray observations of PKS\,2155-304 using \emph{XMM-Newton}, 
spanning over a period of 12 years, are better represented by a broken log parabola function with 
minimal curvature. The index differences are consistent with the spectral 
turnover introduced by synchrotron loss of a parent log parabola electron distribution. 
This study, thereby resolves the inadequacy of the broken 
power-law representation of blazar spectra to explain the index difference under the synchrotron loss phenomena. 
Further, we show that the curvature of the fitted 
function can indicate the energy dependence of the electron escape rate from the main acceleration region.
The simplistic physical model developed under this scenario can fit the observed optical/UV and X-ray data
very well. 

The spectral and temporal properties of PKS\,2155-304 were also studied by \cite{2014MNRAS.444.1077K} using 
\emph{Swift} observations during 2005-2012. The X-ray spectra showed significant curvature and were reproduced better by a log parabola function. The observed anti-correlation between the spectral index at 1 keV and 0.3-10 keV flux suggested that the spectra hardens at high flux states. The non-availability of information at low energy prevents the authors from precise estimation of the peak energy. A log parabola X-ray spectral shape of PKS\,2155-304 was already identified in \emph{XMM-Newton} observations during a period overlapping with the one considered here \citep{2017ApJ...850..209G}. However, the main attempt of these works was to highlight the curved X-ray spectra and their behaviour during different flux states. On the other hand, here we perform a joint spectral fitting of optical--X-ray spectra.
The inadequacy of a log parabola function to explain the combined optical--X-ray spectra was initially shown by \cite{2004A&A...413..489M}, for the case of MKN\,421. In case of PKS\,2155-304, a satisfactory fit of optical--X-ray spectra can be obtained by PLLP model \citep{2014MNRAS.444.3647B}. A plausible scenario under which such a spectrum can be obtained is when the electron escape timescale in ER is energy dependent; whereas, the acceleration and escape timescales in AR are energy independent \citep{2017ApJ...836...83S}. The resultant particle distribution in AR will then be a power-law. On subsequent injection into ER, it will develop a smooth curvature at high energy, imitating a log parabola. On the other hand, here we show that a broken log parabola can provide a better fit without invoking additional number of parameters. This asserts the energy dependence of the escape timescale in AR rather than the ER.

 Through the present work, we demonstrate that the slight curvature observed in addition to the power-law component in the blazar spectra can be tanslated into energy dependence of the escape timescale from the main acceleration region. This identification in turn, can provide clues on the electron diffusion processes in blazar jets. For instance, if the electron diffusion is mainly governed through pitch angle scattering, then this result can be helpful in understanding the magneto hydrodynamic nature of the blazar jets \citep{2012ApJ...745...63S}. 
Alternatively, the information about the energy dependence of the escape timescale can indicate the magnetic field alignment in blazar emission zone \citep{1994A&A...285..687A}. This energy dependence can be coupled with cross-field and/or align-field diffusion coefficients and supplemented with the polarization information, this can be used to draw a global picture regarding the magnetic field structure in blazars.

This research is based on the observations obtained with XMM-Newton satellite, an ESA science mission with instruments and contributions directly funded by ESA member states and NASA. This work has made use of the NASA/IPAC Extra Galactic Database operated by Jet Propulsion Laboratory, California Institute of Technology and the High Energy Astrophysics Science Archive Research Center (HEASARC) provided by 
NASA's Goddard Space Flight Center. JKS thanks Jithesh V and Savithri H Ezhikode for useful discussions. JKS thanks the University Grants Commission for the financial support through the RGNF scheme. Authors thank the anonymous referee for the comments that improved the quality of the manuscript.


\bibliographystyle{mnras}
\bibliography{references} 

\begin{thebibliography}{}
\makeatletter
\relax
\def\mn@urlcharsother{\let\do\@makeother \do\$\do\&\do\#\do\^\do\_\do\%\do\~}
\def\mn@doi{\begingroup\mn@urlcharsother \@ifnextchar [ {\mn@doi@}
  {\mn@doi@[]}}
\def\mn@doi@[#1]#2{\def\@tempa{#1}\ifx\@tempa\@empty \href
  {http://dx.doi.org/#2} {doi:#2}\else \href {http://dx.doi.org/#2} {#1}\fi
  \endgroup}
\def\mn@eprint#1#2{\mn@eprint@#1:#2::\@nil}
\def\mn@eprint@arXiv#1{\href {http://arxiv.org/abs/#1} {{\tt arXiv:#1}}}
\def\mn@eprint@dblp#1{\href {http://dblp.uni-trier.de/rec/bibtex/#1.xml}
  {dblp:#1}}
\def\mn@eprint@#1:#2:#3:#4\@nil{\def\@tempa {#1}\def\@tempb {#2}\def\@tempc
  {#3}\ifx \@tempc \@empty \let \@tempc \@tempb \let \@tempb \@tempa \fi \ifx
  \@tempb \@empty \def\@tempb {arXiv}\fi \@ifundefined
  {mn@eprint@\@tempb}{\@tempb:\@tempc}{\expandafter \expandafter \csname
  mn@eprint@\@tempb\endcsname \expandafter{\@tempc}}}

\bibitem[\protect\citeauthoryear{{Achterberg} \& {Ball}}{{Achterberg} \&
  {Ball}}{1994}]{1994A&A...285..687A}
{Achterberg} A.,  {Ball} L.,  1994, \aap, \href
  {http://adsabs.harvard.edu/abs/1994A%26A...285..687A} {285, 687}

\bibitem[\protect\citeauthoryear{{Aharonian} et~al.,}{{Aharonian}
  et~al.}{2005}]{2005A&A...442..895A}
{Aharonian} F.,  et~al., 2005, \mn@doi [\aap] {10.1051/0004-6361:20053353},
  \href {http://adsabs.harvard.edu/abs/2005A%26A...442..895A} {442, 895}

\bibitem[\protect\citeauthoryear{{Arnaud}}{{Arnaud}}{1996}]{1996ASPC..101...17A}
{Arnaud} K.~A.,  1996, in {Jacoby} G.~H.,  {Barnes} J.,  eds,  Astronomical
  Society of the Pacific Conference Series Vol. 101, Astronomical Data Analysis
  Software and Systems V. p.~17

\bibitem[\protect\citeauthoryear{{Bhagwan}, {Gupta}, {Papadakis}  \&
  {Wiita}}{{Bhagwan} et~al.}{2014}]{2014MNRAS.444.3647B}
{Bhagwan} J.,  {Gupta} A.~C.,  {Papadakis} I.~E.,   {Wiita} P.~J.,  2014,
  \mn@doi [\mnras] {10.1093/mnras/stu1703}, \href
  {http://adsabs.harvard.edu/abs/2014MNRAS.444.3647B} {444, 3647}

\bibitem[\protect\citeauthoryear{{Chiappetti} et~al.,}{{Chiappetti}
  et~al.}{1999}]{1999ApJ...521..552C}
{Chiappetti} L.,  et~al., 1999, \mn@doi [\apj] {10.1086/307577}, \href
  {http://adsabs.harvard.edu/abs/1999ApJ...521..552C} {521, 552}

\bibitem[\protect\citeauthoryear{{Gaur}, {Chen}, {Misra}, {Sahayanathan}, {Gu},
  {Kushwaha}  \& {Dewangan}}{{Gaur} et~al.}{2017}]{2017ApJ...850..209G}
{Gaur} H.,  {Chen} L.,  {Misra} R.,  {Sahayanathan} S.,  {Gu} M.~F.,
  {Kushwaha} P.,   {Dewangan} G.~C.,  2017, \mn@doi [\apj]
  {10.3847/1538-4357/aa95bc}, \href
  {http://adsabs.harvard.edu/abs/2017ApJ...850..209G} {850, 209}

\bibitem[\protect\citeauthoryear{{Heavens} \& {Meisenheimer}}{{Heavens} \&
  {Meisenheimer}}{1987}]{1987MNRAS.225..335H}
{Heavens} A.~F.,  {Meisenheimer} K.,  1987, \mn@doi [\mnras]
  {10.1093/mnras/225.2.335}, \href
  {http://adsabs.harvard.edu/abs/1987MNRAS.225..335H} {225, 335}

\bibitem[\protect\citeauthoryear{{Kapanadze}, {Romano}, {Vercellone}  \&
  {Kapanadze}}{{Kapanadze} et~al.}{2014}]{2014MNRAS.444.1077K}
{Kapanadze} B.,  {Romano} P.,  {Vercellone} S.,   {Kapanadze} S.,  2014,
  \mn@doi [\mnras] {10.1093/mnras/stu1504}, \href
  {http://adsabs.harvard.edu/abs/2014MNRAS.444.1077K} {444, 1077}

\bibitem[\protect\citeauthoryear{{Kardashev}}{{Kardashev}}{1962}]{1962SvA.....6..317K}
{Kardashev} N.~S.,  1962, \sovast, \href
  {http://adsabs.harvard.edu/abs/1962SvA.....6..317K} {6, 317}

\bibitem[\protect\citeauthoryear{{Madejski} et~al.,}{{Madejski}
  et~al.}{2016}]{2016ApJ...831..142M}
{Madejski} G.~M.,  et~al., 2016, \mn@doi [\apj] {10.3847/0004-637X/831/2/142},
  \href {http://adsabs.harvard.edu/abs/2016ApJ...831..142M} {831, 142}

\bibitem[\protect\citeauthoryear{{Mankuzhiyil}, {Persic}  \&
  {Tavecchio}}{{Mankuzhiyil} et~al.}{2010}]{2010ApJ...715L..16M}
{Mankuzhiyil} N.,  {Persic} M.,   {Tavecchio} F.,  2010, \mn@doi [\apjl]
  {10.1088/2041-8205/715/1/L16}, \href
  {http://adsabs.harvard.edu/abs/2010ApJ...715L..16M} {715, L16}

\bibitem[\protect\citeauthoryear{{Mankuzhiyil}, {Ansoldi}, {Persic}, {Rivers},
  {Rothschild}  \& {Tavecchio}}{{Mankuzhiyil}
  et~al.}{2012}]{2012ApJ...753..154M}
{Mankuzhiyil} N.,  {Ansoldi} S.,  {Persic} M.,  {Rivers} E.,  {Rothschild} R.,
   {Tavecchio} F.,  2012, \mn@doi [\apj] {10.1088/0004-637X/753/2/154}, \href
  {http://adsabs.harvard.edu/abs/2012ApJ...753..154M} {753, 154}

\bibitem[\protect\citeauthoryear{{Massaro}, {Perri}, {Giommi}  \&
  {Nesci}}{{Massaro} et~al.}{2004}]{2004A&A...413..489M}
{Massaro} E.,  {Perri} M.,  {Giommi} P.,   {Nesci} R.,  2004, \mn@doi [\aap]
  {10.1051/0004-6361:20031558}, \href
  {http://adsabs.harvard.edu/abs/2004A%26A...413..489M} {413, 489}

\bibitem[\protect\citeauthoryear{{Massaro}, {Tramacere}, {Cavaliere}, {Perri}
  \& {Giommi}}{{Massaro} et~al.}{2008}]{2008A&A...478..395M}
{Massaro} F.,  {Tramacere} A.,  {Cavaliere} A.,  {Perri} M.,   {Giommi} P.,
  2008, \mn@doi [\aap] {10.1051/0004-6361:20078639}, \href
  {http://adsabs.harvard.edu/abs/2008A%26A...478..395M} {478, 395}

\bibitem[\protect\citeauthoryear{{Rieger}, {Bosch-Ramon}  \& {Duffy}}{{Rieger}
  et~al.}{2007}]{2007Ap&SS.309..119R}
{Rieger} F.~M.,  {Bosch-Ramon} V.,   {Duffy} P.,  2007, \mn@doi [\apss]
  {10.1007/s10509-007-9466-z}, \href
  {http://adsabs.harvard.edu/abs/2007Ap%26SS.309..119R} {309, 119}

\bibitem[\protect\citeauthoryear{Sahayanathan, Sinha  \& Misra}{Sahayanathan
  et~al.}{2018}]{RAA4107}
Sahayanathan S.,  Sinha A.,   Misra R.,  2018, \mn@doi [RAA]
  {10.1088/1674–4527/18/3/35}, 18, 35

\bibitem[\protect\citeauthoryear{{Sambruna}, {Barr}, {Giommi}, {Maraschi},
  {Tagliaferri}  \& {Treves}}{{Sambruna} et~al.}{1994}]{1994ApJS...95..371S}
{Sambruna} R.~M.,  {Barr} P.,  {Giommi} P.,  {Maraschi} L.,  {Tagliaferri} G.,
   {Treves} A.,  1994, \mn@doi [\apjs] {10.1086/192102}, \href
  {http://adsabs.harvard.edu/abs/1994ApJS...95..371S} {95, 371}

\bibitem[\protect\citeauthoryear{{Schlafly} \& {Finkbeiner}}{{Schlafly} \&
  {Finkbeiner}}{2011}]{2011ApJ...737..103S}
{Schlafly} E.~F.,  {Finkbeiner} D.~P.,  2011, \mn@doi [\apj]
  {10.1088/0004-637X/737/2/103}, \href
  {http://adsabs.harvard.edu/abs/2011ApJ...737..103S} {737, 103}

\bibitem[\protect\citeauthoryear{{Seaton}}{{Seaton}}{1979}]{1979MNRAS.187P..73S}
{Seaton} M.~J.,  1979, \mn@doi [\mnras] {10.1093/mnras/187.1.73P}, \href
  {http://adsabs.harvard.edu/abs/1979MNRAS.187P..73S} {187, 73P}

\bibitem[\protect\citeauthoryear{{Sinha}, {Sahayanathan}, {Acharya}, {Anupama},
  {Chitnis}  \& {Singh}}{{Sinha} et~al.}{2017}]{2017ApJ...836...83S}
{Sinha} A.,  {Sahayanathan} S.,  {Acharya} B.~S.,  {Anupama} G.~C.,  {Chitnis}
  V.~R.,   {Singh} B.~B.,  2017, \mn@doi [\apj] {10.3847/1538-4357/836/1/83},
  \href {http://adsabs.harvard.edu/abs/2017ApJ...836...83S} {836, 83}

\bibitem[\protect\citeauthoryear{{Summerlin} \& {Baring}}{{Summerlin} \&
  {Baring}}{2012}]{2012ApJ...745...63S}
{Summerlin} E.~J.,  {Baring} M.~G.,  2012, \mn@doi [\apj]
  {10.1088/0004-637X/745/1/63}, \href
  {http://adsabs.harvard.edu/abs/2012ApJ...745...63S} {745, 63}

\makeatother
\end{thebibliography}

\appendix
\section{Log Parabola Electron Distribution undergoing radiative losses} \label{app:lp}
A log parabola electron distribution, $u(\gamma)$, undergoing synchrotron loss can be described by
\begin{align}
\label{eq:app1}
	-\frac{d}{d\gamma} \left[\zeta \gamma^2 u(\gamma)\right] = Q_0\, \gamma^{-a-b \,\rm{ln}\, \gamma}
\end{align}
Here, $\zeta \gamma^2$ is the radiative energy loss rate. Equation (\ref{eq:app1}) can be rewritten
as
\begin{align}
\label{eq:app2}
\frac{d\,\rm{ln}\,W(\gamma)}{d\,\rm{ln}\,\gamma} =-\frac{Q_0}{\zeta W(\gamma)}\gamma^{-a+1-b \,\rm{ln}\, \gamma}
\end{align}
where, $W(\gamma)=\gamma^2\, u(\gamma)$. We assume {a priori} that the solution $ W(\gamma)$ is of a log-parabola type:
\begin{align}\label{eq:app3}
 W(\gamma)= W_0\, \gamma^{-c-d\,\rm{ln}\,\gamma}
 \end{align}
and recast equation (\ref{eq:app2}) to
\begin{align}\label{eq:app4}
	\frac{d\,\rm{ln}\,W(\gamma)}{d\,\rm{ln}\gamma} =  -\frac{Q_0}{\zeta W_0} \gamma^\kappa \approx-\frac{Q_0}{\zeta W_0}(1+\kappa\,\rm{ln}\,\gamma)
\end{align}
where, $\kappa=-a+c+1-(b-d)\,\rm{ln}\,\gamma\,\ll \,1$. Using equation (\ref{eq:app3})
\begin{align}
\label{eq:app5}
\frac{d\,\rm{ln}\,W(\gamma)}{d\,\rm{ln}\,\gamma} = -c-2d\,\rm{ln}\,\gamma
\end{align}
Comparing equations (\ref{eq:app4}) and (\ref{eq:app5}), we note that for consistency $\kappa$ is required
to be energy independent and hence, $d=b$. Moreover, $c(c-a+1) = 2b$ and for $b\ll 1$,
\begin{align*}
	c \approx (a-1)\left[1+\frac{2b}{(a-1)^2}\right]
\end{align*} 
The resultant electron distribution will be,
\begin{align}
	u(\gamma)\propto \gamma^{-(a+1)-b\,\rm{ln}\, \gamma - \frac{2b}{a-1}}
\end{align}
Hence, the synchrotron loss process steepens the injected log parabola distribution by $1+\frac{2b}{a-1}$.

\noindent

\bsp	
\label{lastpage}
\end{document}